\documentclass[%
prb,twocolumn,showpacs,
amssymb,amsmath%
]{revtex4-1}

\usepackage{ifpdf}
\ifx\pdfoutput\undefined
	\pdffalse              	
\else
	\pdfoutput=1           
	\pdftrue               	
\fi

\ifpdf
\usepackage[activate]{pdfcprot}				
\else											
\fi

\ifpdf 
	\usepackage[pdftex]{graphicx} 
\else
	\usepackage[dvips]{graphicx} 
\fi

\ifpdf
	\DeclareGraphicsExtensions{.pdf,.jpg,.png}
\else
	\DeclareGraphicsExtensions{.eps}
\fi

\usepackage{hypernat}
\usepackage{color}
\usepackage{hyperref}
\hypersetup{colorlinks=true, linkcolor=blue, urlcolor=red}

\usepackage{natbib}
\usepackage{amsfonts}%
\usepackage{units}
\usepackage{nicefrac}
\usepackage{setspace}
\usepackage{threeparttable}
\usepackage[vflt]{floatflt}
\usepackage{dcolumn}
\usepackage{bbm}
\usepackage{bm}
\usepackage{braket}

\graphicspath{{figures/}}

\begin{document}
\title{The electron spectrum of epitaxial graphene monolayers}

\date{\today} 

\author{O. Pankratov}\email[email: ]{oleg.pankratov@physik.uni-erlangen.de}

\author{S. Hensel}

\author{M. Bockstedte}

\affiliation{Lehrstuhl f\"ur Theoretische Festk\"orperphysik,
  Universit\"at Erlangen-N\"urnberg, Staudtstrasse 7 B2, D-91058
  Erlangen, Germany}


\begin{abstract}
  Epitaxial graphene on SiC possesses, quite remarkably, an electron spectrum
  similar to that of freestanding samples. Yet, the coupling to the substrate,
  albeit small, affects the quasiparticle properties. Combining \emph{ab
    initio} calculations with symmetry analysis, we derive a modified
  Dirac-Weyl Hamiltonian for graphene epilayers. While for the epilayer on the
  C-face the Dirac cone remains almost intact, for epilayers on the Si-face
  the band splitting is about 30\,meV. At certain energies, the Dirac bands
  are significantly distorted by the resonant interaction with interface
  states, which should lead to mobility suppression, especially on the
  Si-face.
\end{abstract}
\maketitle

\definecolor{Greener}{rgb}{.0,.5196,.004}
\definecolor{darkBlue}{rgb}{.004,.0,.50196}

Continuing improvement of epitaxial graphene on SiC \cite{Seyller08,Hass08}
that culminated in observations of the quantum Hall effect \cite{Wu09,Jobst10}
and of a very high mobility at the Dirac point \cite{Orlita08} have raised the
{ranking of this material} dramatically. Apart from the capability to emulate
freestanding graphene, epitaxial graphene has a number of specific qualities
making it an intersting material in its own right. These features stem from
the interaction with the underlying substrate - the Si or C terminated SiC
surface. Interestingly, the growth of graphene differs drastically on both
surfaces.  On the Si-face the growth is slow thus facilitating the fabrication
of a single monolayer.\cite{Seyller08}  The subsequent C-layers arrange in
the common graphite-type AB (Bernal) stacking.  On the C-face, the much faster
growth typically yields multilayer stacks of mutually rotated C-layers.\cite{Hass08} 
The rotation is very important since it decouples {individual}
layers electronically, such that the whole stack behaves effectively as a
single graphene sheet.\cite{Hass08,Shallcross10} Recently {also monolayer}
graphene has been achieved on the C-face.\cite{Wu09a} Compared with the
C-face, the electron mobility in Si-face graphene is much lower.  Along with
the preference for Bernal stacking of the latter, this indicates a stronger
coupling to the substrate.

In fact, the {graphene monolayer} does not reside directly on SiC but rather
on some buffer layer, as first realized theoretically
\cite{Mattausch07,Varchon07} and then confirmed experimentally.\cite{Seyller08,Riedl09} 
The currently accepted buffer model for the Si-face
is a corrugated graphene layer, that is covalently bonded to the substrate
fitting into a $(6\sqrt{3}\times6\sqrt{3})$R30 surface reconstruction.\cite{Kim08} 
The reconstruction unit cell almost exactly coincides with a
(13$\times$13) graphene unit providing the commensurable base for subsequent
graphene layers. The strong covalent interaction with the substrate completely
erases all Dirac-Weyl features of the buffer.  Hence it is the second carbon
layer, which exhibits the graphene-like band structure and is referred to as
``monolayer graphene''.\cite{Mattausch07,Varchon07} This scenario was
convincingly confirmed in recent experiments with hydrogen intercalation.\cite{Riedl09} 
Diffusing underneath the buffer H atoms cause its release and
as a result, the formation of a quasi-freestanding Bernal-stacked bilayer was
observed.

The situation on the C-face is less clear. A distinct buffer layer has not yet
been identified and it was speculated, that already the very
first carbon sheet might be graphene-like. However, this contradicts
calculations \cite{Mattausch07,Varchon07} which clearly show the extinction
of the Dirac spectrum of the first C-layer on both SiC surfaces. Si adatoms
\cite{Magaud09} or a corrugated C-layer \cite{Mattausch07,Varchon08} were
considered as buffer models on the C-face. Generally, the much weaker
graphene-substrate coupling on a C-face suggests, that in this
case the particular interface structure is not as important as on the Si-face.

The graphene-substrate coupling, and especially its effect on the electron
spectrum in the vicinity of the Dirac point, is of particular interest for
electron transport. It has been a subject of a long debate whether the
graphene-substrate interaction opens an energy gap with experimental
estimates ranging from zero to \unit[0.3]{eV}.\cite{Seyller08,Zhou07,Vitali08}

\begin{figure}[ht]
\includegraphics[width=\linewidth]{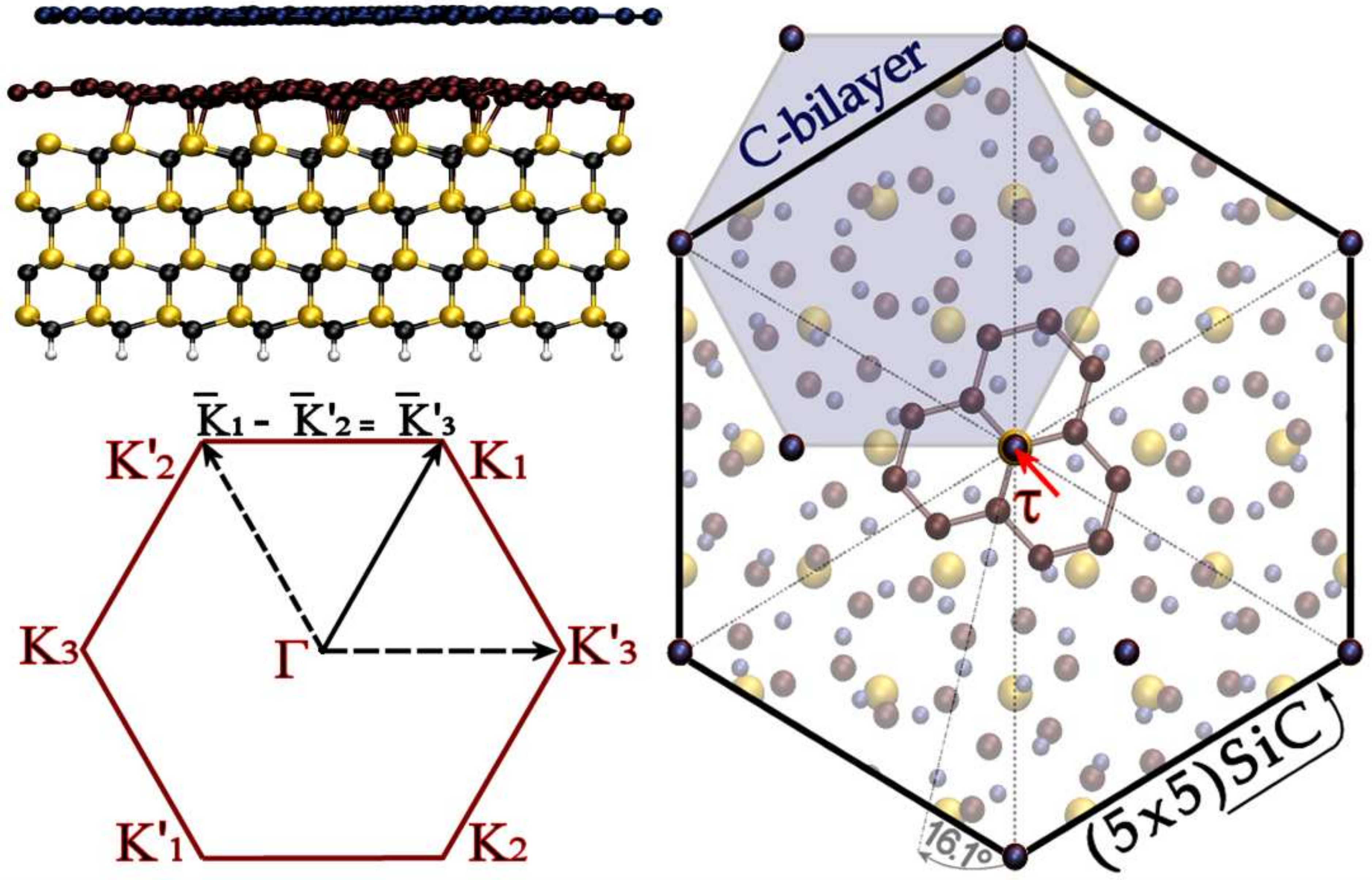}
\caption{\label{fig:geometry}(Color online) Top and side views of the rotated
  graphene monolayer on a (5$\times$5) SiC$(0001)$/buffer slab with the
  H-passivated bottom (C/Si atoms: small dark/large light spheres).
  The shaded area indicates the ($\sqrt{13}$$\times$$\sqrt{13}$)-graphene
  Wigner-Seitz cell of the graphene/buffer bilayer (medium-sized/small
  spheres) with relative rotation of $\unit[32.2]{^\circ}$, that is
  commensurate with the (5$\times$5) SiC cell (surface atoms: large spheres)
  -- the three central hexagons of the graphene layer are emphasized along with
  the vector $\bm{\tau}$ (cf.\ text).  Lower left: The Brillouin zone
  of an unperturbed graphene layer.}
\end{figure}

In this article we consider the electronic structure of graphene monolayers on
SiC combining \emph{ab initio} calculations with symmetry analysis. We derive
the low energy Hamiltonian, that replaces the Dirac-Weyl Hamiltonian of a
freestanding graphene. Instead of the computationally demanding
$(6\sqrt{3}\times6\sqrt{3})$ structure, we adopt a strain free,
commensurable (5$\times$5) interface model with a corrugated carbon layer as a
buffer \mbox{(cf.\ Fig.\ \ref{fig:geometry})}.  The (5$\times$5) reconstruction is
observed, although more rarely, on the Si-face.\cite{Riedl07} With the same
buffer model for the C-face we find a much weaker graphene-SiC coupling than
on the Si-face.

Backfolding of the graphene K and K$'$ points to the $\Gamma$-point in the
(5$\times$5) Brillouin zone produces four closely lying energy branches. On
the Si-face, we find that the Dirac cone is split by about \unit[30]{meV},
whereas on the C-face two branches of an essentially unperturbed cone exist
and the other two are separated by a very small gap (\unit[$<10$]{meV}). The
splitting of the Dirac cones \emph{is not} due to the corrugation of the
graphene layer, but arises from the interaction with the substrate. We
verified this by calculating the freestanding carbon layer with the same
atomic positions, for which we found a perfect Dirac cone.

The buffer layer is commensurate with the SiC surface when rotated by
$\sim\pm\unit[16.1]{^\circ}$, such that the resulting structure possesses a
(5$\times$5) periodicity relative to the SiC substrate \mbox{(cf.\ Fig.\
  \ref{fig:geometry})}. The following graphene epilayer can be either rotated
by $\sim\mp\unit[32.2]{^\circ}$ with respect to the buffer or aligned in
AB-stacking. As seen in the right panel of \mbox{Fig.\ \ref{fig:geometry}},
the (5$\times$5) structure naturally results from a commensuration of the
twisted carbon bilayer and the SiC surface. The (5$\times$5) SiC unit cell is
almost perfectly commensurate with the 
($\sqrt{13}$$\times$$\sqrt{13}$) graphene cell. Notably, the rotation angle of $\unit[(30\pm
2.2)]{^\circ}$ dominates in the multilayered graphene stacks \cite{Hass08} and a $\unit[30]{^\circ}$-rotation of the graphene bilayer
relative to the SiC cell is a common feature for both Si- and C-terminated
surfaces.\cite{Seyller08,Hass08}


We used the VASP density functional package\cite{Kresse96} to obtain the
interface geometry by relaxing atomic positions in the (5$\times$5) unit cell
and to calculate the band structures.\footnote{The calculations utilized a
  plane wave basis (\unit[358]{eV} cut-off), ultrasoft pseudopotentials
  [D. Vanderbilt, Phys. Rev. B 41, 7892 (1990)], LDA approximation
  [D. M. Ceperley and B. J. Alder, Pys. Rev. Lett. 45, 566 (1980).] and {a}
  $\Gamma$-centered 7$\times$7$\times$1 k-point grid. All structures were
  relaxed until forces amounted less than $10^{-2}$\,eV/$\mathrm{\AA}$ keeping the
  bottom SiC bilayer at bulk position.} The buffer layer shows a significant
corrugation \mbox{(cf.\ Fig.\ \ref{fig:geometry})}, that is partly
transmitted into the top C-layer.
\begin{figure*}[t]
\includegraphics[width=1.0\linewidth]{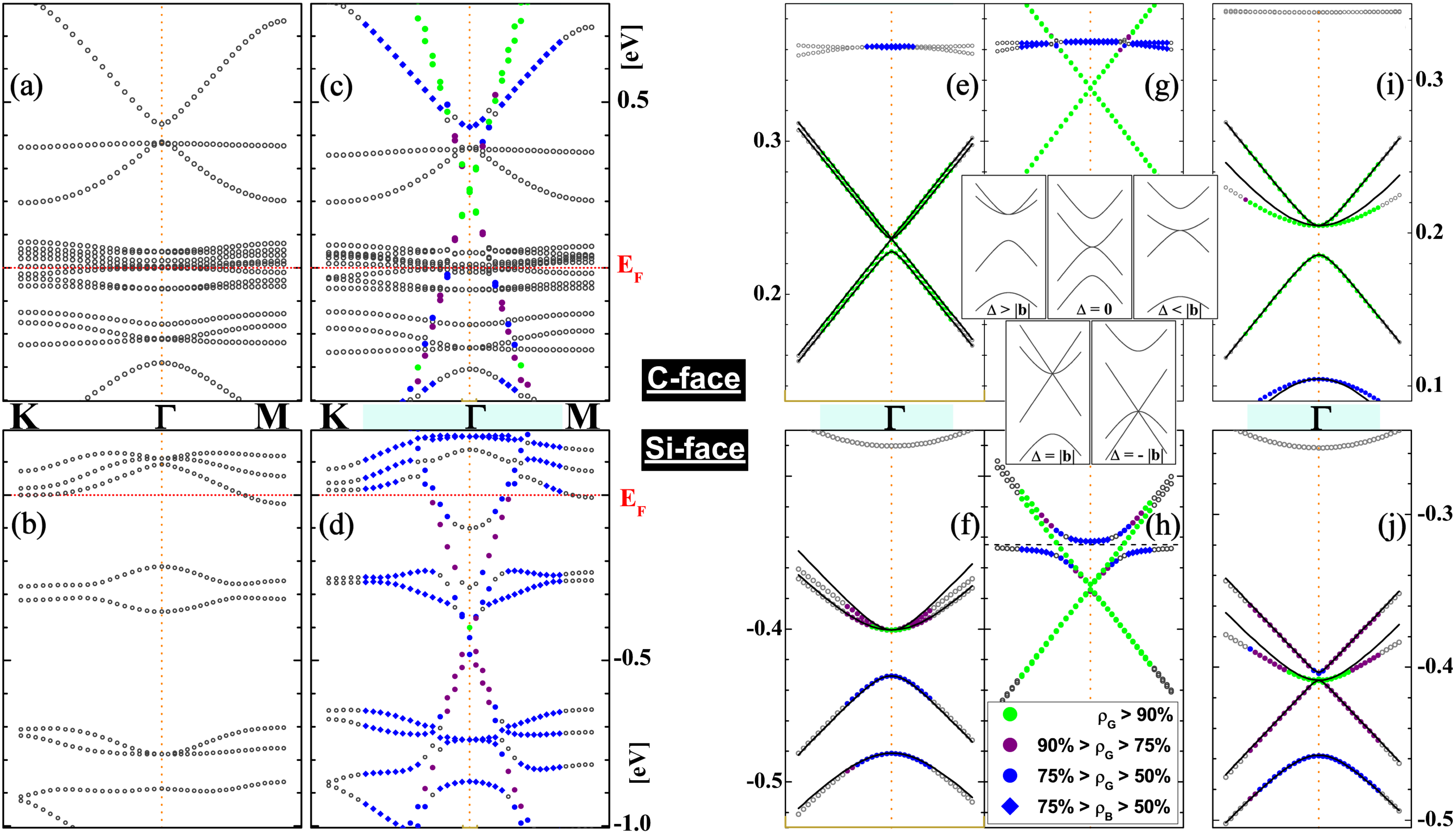}
\caption{\label{fig:electronics}(Color online) Calculated band structure of
  epitaxial graphene on SiC$(000\bar{1})$ (C-face, upper row) and on
  SiC$(0001)$ (Si-face, lower row). Large-scale energy spectra (a),(b) of the
  buffer on SiC and (c),(d) of the $\unit[32.2]{^\circ}$ rotated
  buffer/epilayer structure. The high resolution spectra in the range
  $[\nicefrac{1}{20}\ \overline{\text{K}\Gamma},\nicefrac{1}{20}\
  \overline{\Gamma\text{M}}]$ are shown for (e)-(h): the rotated
  buffer/epilayer with the equilibrium buffer-epilayer distance [(e),(f)] and with a distance artificially increased by \unit[1]{\AA}
  [(g),(h)]; Bernal stacking [(i),(j)]. The color code indicates the degree of
  localization of states on the graphene epilayer ($\rho_\text{G}$) and on the
  buffer layer ($\rho_\text{B}$) as deduced from the integrated density
  $\varrho(z) = \int |\Psi|^2\,dx\,dy$. Fitted bands are shown as solid curves
  \mbox{(cf.\ Eq.\ (\ref{Eigenvalue})} and \mbox{Tab.~\ref{tab:fits}}). The insets
  illustrate the possible types of band structures that follow from \mbox{Eq.\
    (\ref{Eigenvalue})}.}
\end{figure*}
The calculated large-scale band structure is depicted in \mbox{Fig.\
  \ref{fig:electronics}(a)-(d)}. The graphene-type linear bands appear with
the second carbon layer \mbox{[Fig.\
  \ref{fig:electronics}(c),(d)]}. The
buffer does not possess that feature, but supplies a number of flat interface
states, which pin the Fermi level E$_\text{F}$ \mbox{[Fig.\
  \ref{fig:electronics}(a),(b)]}. Owing to Fermi level pinning, the epilayer is
either n- or p-doped on the Si- and C-face respectively, in accord with the
measurements. \cite{Seyller08,Hass08,Miller09} It is also visible, that the
resonant interaction with the interface states causes a significant distortion
of the Dirac bands.

\mbox{Figures\ \ref{fig:electronics}(e)-(j)} show the high-resolution energy
spectra close to the Dirac point. Regardless whether the twisted or the
Bernal-type structure is chosen, the energy spectrum resembles a modified
\emph{single layer} spectrum (with folded K and K$'$ points) and \emph{not} a
graphene bilayer spectrum. This again confirms, that the buffer does not
possess the Dirac states at the K-point, i.e.\ it is completely passivated by
the substrate. The substrate potential is rather weak, especially on the
C-face, causing the band splittings in the range of $\unit[5-30]{meV}$. As
pointed out above, the slight warping of the epilayer does not affect its band
structure within an accuracy of \unit[2]{meV}. Hence it should be possible to
obtain the low-energy spectra by accounting for the substrate as a
perturbation of the ideal graphene states.  In the following we use symmetry
considerations to construct the effective Hamiltonian for the graphene
monolayer. The derived Hamiltonian reproduces the calculated \emph{ab initio}
  energy spectra and can serve for the description of the quasiparticle
dynamics.

Commonly, the graphene spectrum is introduced via the tight-binding modeling
of the $\pi$-bands. \cite{Neto09} Yet, the symmetry underlying the spectrum is
much more transparently expressed within the empty lattice approach.  The
three states $\ket{\mathbf{K}_i}$ ($i= 1,2,3$), that correspond to the plane
waves $\varphi_{K_i}(\bm{r})$ at equivalent corners $K_i$ of the Brillouin
zone, combine into the K-point eigenstates and similarly for the time-reversed
states at points $K'_{i}$.  The trigonal symmetry dictates the Hamiltonian
matrix
\begin{eqnarray} \label{Hamilton}
 \widehat{H}_K &=& \hspace{5ex}
   \left(
     \begin{array}{ccc}
       \frac{v_\text{F}}{3}\bm{n}_1\bm{k} 	& V\color{white}{*} & V^*\\
       V^* & \frac{v_\text{F}}{3}\bm{n}_2\bm{k}& V\color{white}{*}\\
       V\color{white}{*}& V^*&\frac{v_\text{F}}{3}\bm{n}_3\bm{k}	
   \end{array}
 \right)\nonumber\\ 
 &=& \frac{v_\text{F}}{3} \text{diag}\left(\bm{n}_i\cdot \bm{k} \right) + 
         2|V|\cos{\left(\frac{2\pi}{3} \widehat{L}_z + \varphi\right)},
\end{eqnarray}
where the mixing of the basis states is expressed via the symmetrized operator
of a $(\nicefrac{2\pi}{3})$-rotation: $\widehat{c}_3 + \widehat{c}^\dagger_3 =
\exp{(i \frac{2\pi}{3} \widehat{L}_z)} + h.c.$, with the angular momentum
operator $\widehat{L}_z$ in the Hilbert space $l=1$. The diagonal elements in
\mbox{{Eq.}~(\ref{Hamilton})} are linear invariants of a small displacement $\bm{k}$
from the K-point and the unit vectors $\bm{n}_i =
{\mathbf{K}_i}/{|\mathbf{K}_i|}$. The phase $\varphi$ of the matrix elements
$V = |V|\, \exp{(i\varphi)}$ depends on the choice of the coordinate
origin. For ideal graphene $\varphi$ takes the values $0$ or $\pm
\nicefrac{2\pi}{3}$. In an epitaxial layer, however, the inversion symmetry
and hence the mirror planes of the small group $C_{3v}$ are lost. This implies
a reduction to $C_3$ and allows arbitrary values of $\varphi$. By time
  reversal the Hamiltonian at the $K'$-point is
\begin{equation} \label{Hamilton'}
 \widehat{H}_{K'} = -\frac{v_\text{F}}{3} \text{diag}\left(\bm{n}_{i}\cdot \bm{k}
 \right) 
 + 2|V|\cos{\left(\frac{2\pi}{3} \widehat{L}_z - \varphi\right)}.
\end{equation}
Replacing $\widehat{L}_z$ by its eigenvalues ($0,\pm 1$) we obtain the energy
levels at the K-point
\begin{eqnarray} \label{Energy}
 \varepsilon = 2|V|\,\cos{\varphi} \quad\text{and}\quad \varepsilon_{\pm} 
= 2|V|\,\cos{\left(\frac{2\pi}{3}\pm\varphi\right)}.
\end{eqnarray}
Of special interest are the two states, that form the tip of the Dirac cone
in ideal graphene. Depending on $\varphi$ these can be any two of the levels
in \mbox{Eq.~(\ref{Energy})}. The choice $\varphi = 0$ for the coordinate origin in the
center of the graphene hexagon selects $\varepsilon_+$ and $\varepsilon_-$.

The states $\ket{\mathbf{K}_i}$ and $\ket{\mathbf{K}'_{i}}$ are coupled by the
Umklapp process due to the substrate potential $V(\bm{r})$. Since we account
for $V(\bm{r})$ perturbatively, we can assume that it preserves (within the
intervalley matrix elements) point symmetry of a pristine surface. However,
the point group centers of the epilayer and of the substrate (i.e.\ of the
last Si or C atomic layer) are displaced by the vector $\bm{\tau}$, which
connects two adjacent graphene atoms \mbox{(cf. Fig.~\ref{fig:geometry})}. This displacement generates
a phase factor of the intervalley matrix elements
\begin{eqnarray} \label{Coupling}
 V_{ij} &=& \int \varphi^*_{K_i}(\bm{r}) \, V(\bm{r}+\bm{\tau})\,\varphi_{K'_j}(\bm{r})\, d\bm{r}\nonumber \\
        &=& e^{-i\left(\bm{K}_i-\bm{K}'_{j}\right)\bm{\tau}} \int \varphi^*_{K_i}(\bm{r}) \, V(\bm{r})\,\varphi_{K'_i}(\bm{r})\, d\bm{r}. \label{Coupling_1}
\end{eqnarray}
The shift of the coordinate origin in \mbox{Eq.~(\ref{Coupling})} enables us to exploit
the trigonal symmetry, which requires $|{V}_{13}| =
|{V}_{21}| = |{V}_{32}|$.

Given $\bm{\tau}$ and $\bm{K}_i - \bm{K}'_{j}$ \mbox{(cf.\ Fig.\
  \ref{fig:geometry})} one finds, that the phase in \mbox{Eq.~(\ref{Coupling_1})}
takes values $\pm\nicefrac{2\pi}{3}$ or 0. For $V_{ij}$ and $V_{ji}$ the phase
factors can be made complex conjugated by shifting phases of the basis
K-functions: $\ket{\mathbf{K}_1} \rightarrow
e^{i\frac{2\pi}{3}}\ket{\mathbf{K}_1}$, $\ket{\mathbf{K}_2} \rightarrow
\ket{\mathbf{K}_2}$, $\ket{\mathbf{K}_3} \rightarrow
e^{-i\frac{2\pi}{3}}\ket{\mathbf{K}_3}$.  This, however, changes the phases in
\mbox{{Eq.}~(\ref{Hamilton})}, such that the K-Hamiltonian takes the form
\begin{eqnarray} \label{Hamilton_phase} \widehat{H}_K = \frac{v_\text{F}}{3}
  \text{diag}\left(\bm{n}_i\cdot \bm{k} \right) +
  2|V|\cos{\left(\frac{2\pi}{3} \widehat{L}_z -\frac{2\pi}{3}+
      \varphi\right)}.
\end{eqnarray}
After this phase transformation the intervalley interaction \mbox{Eq.~(\ref{Coupling})}
acquires a structure similar to \mbox{Eqs.~(\ref{Hamilton'})} and (\ref{Hamilton_phase}):
\begin{eqnarray}\label{scatteringOP}
  \widehat{V} = \alpha + 2\/\beta \cos{\left( \frac{2\pi}{3} \widehat{L}_z 
 + \frac{2\pi}{3}\right)} \quad (\alpha, \beta \in \mathbbm{C}).
\end{eqnarray}
In the diagonal representation ($\widehat{L}_z = diag(-1, 0, 1)$) the Hamiltonian matrix reads
\begin{eqnarray}\label{H-matrix}
   \widehat{H} = \left(
    \begin{array}{cccccc}
     \varepsilon_{+} 	& p^*& p\color{white}{*} & \alpha + 2\beta & 0& 0\\
                        & \varepsilon_{-}& p^*& 0\,\,\,& \alpha - \beta& 0\\
                   && \varepsilon\color{white}{*}\, & 0\,\,\,& 0& \alpha - \beta\\
                   &&&\color{white}{-}\color{black}\varepsilon_{+}\color{white}{*}& -p^*& -p\color{white}{*}\\
             & c.c.&&& \varepsilon& -p^*\\
                   &&&&& \color{white}{-}\color{black}\varepsilon_{-}		      	
    \end{array}
  \right),
\end{eqnarray}
where \mbox{$p = v_\text{F}(k_x + ik_y)$} with the Fermi velocity $v_\text{F}$. Note, that
the phase shift in \mbox{Eq.~(\ref{Hamilton_phase})} leads to a cyclic permutation of the
K-point eigenvalues. The ``Dirac part'' $\widehat{H}_D$ of \mbox{Eq.~(\ref{H-matrix})}
comprises the energy levels, which merge for $\varphi = 0$. Introducing
parameters $\Delta = 2\sqrt{3}|V|\sin{\varphi}$ and $b = \nicefrac{\alpha}{2}
+ \beta$ we obtain
\begin{eqnarray}\label{HD-matrix}
   \widehat{H}_D = \left(
    \begin{array}{cccc}
  -\Delta& \,\,p^*& \,\,\,2b& \,0\\
  \,\,\,p& \Delta& \,\,\,0& \,0\\
  \,\,\,\,2b^*&0&-\Delta& -p^*\\
  \,\,0&0&-p& \,\Delta
    \end{array}
  \right).
\end{eqnarray}
The eigenvalues of $\widehat{H}_D$ are 
\begin{eqnarray}\label{Eigenvalue}
  \varepsilon_i(k) = \pm |b| \,(\pm)\, \sqrt{\big(\Delta \mp |b|\big)^2 + \big(v_\text{F}\, k\big)^2}
\end{eqnarray}
with $i=1,\ldots,4$ for different combinations of signs. Equation
(\ref{Eigenvalue}) describes the four bands originating from the K and K$'$
states of an ideal graphene layer subject to a symmetry lowering potential of
the substrate. The spectrum of \mbox{Eq.~(\ref{Eigenvalue})} takes qualitatively
different forms for $\Delta > |b|$, $\Delta \cong |b|$ and $\Delta < |b|$
\mbox{(cf. insets Fig.\ \ref{fig:electronics})}.  By fitting \mbox{Eq.\
  (\ref{Eigenvalue})} to the \emph{ab initio} bands, we determine the
Hamiltonian parameters as given in \mbox{Tab.\ \ref{tab:fits}}. The $\unit[32]{^\circ}$-twisted buffer/epilayer
pair on the Si-face and the Bernal stacking on the C-face matches the $\Delta >
|b|$ case, whereas the twisted C-termination and the Bernal stacking on
the Si-face correspond to the $\Delta \cong |b|$ case.
\begingroup
\renewcommand{\arraystretch}{1.3}
\squeezetable
\begin{table}[!ht]
  \begin{ruledtabular}
  \begin{tabular}{ccccc}
buffer/epilayer&\multicolumn{2}{l}{$\unit[32.2]{^\circ}$-twisted}
                                 &\multicolumn{2}{l}{Bernal-type} \\
\mbox{[meV]}&$|b|$ & $\Delta$&$|b|$ & $\Delta$\\
      \hline
Si-face $(0001)$& 12.7& 21.7& 13.5& 11.0\\
C-face  $(000\bar{1})$&  2.1& 2.1& 20.3	& 29.9\\
  \end{tabular}
  \end{ruledtabular}
  \caption{\label{tab:fits} 
Parameters $\Delta$ and $|b|$ as obtained from a fit of
\mbox{Eq. (\ref{Eigenvalue})} to the \emph{ab initio} band structure. 
The fit yields
$v_{\text{F}}=0.8\,v_{\text{F}}^{\text{free}}$ (Si-face) and
$v_{\text{F}}=0.96\,v_{\text{F}}^{\text{free}}$ (C-face) with
$v_{\text{F}}^{\text{free}}=8.37\cdot10^{5}$\,m\/s$^{-1}$ being the calculated
Fermi velocity of freestanding graphene.}
\end{table} 
\endgroup

The fitted bands of \mbox{Eq.~(\ref{Eigenvalue})} are shown in \mbox{Fig.\
\ref{fig:electronics}(e),(f)} and (j),(i).  
A noticeable deviation of the fit from the \emph{ab initio} data occurring for
some bands originates in the repulsion by the closely lying interface
states (see below).

Setting $\Delta = 0$ in \mbox{Eq.\ (\ref{HD-matrix})}, $\widehat{H}_D$
becomes similar to the Hamiltonian of the freestanding twisted bilayer, where
the spectrum is formed by the K-states of the two graphene layers. Due to the
inequivalence of neighboring atoms, the intervalley
scattering in \mbox{Eq.~(\ref{scatteringOP})} has the same structure as in the case
of the ``odd
sublattice exchange'' of the twisted bilayer.\cite{Mele10}

In graphene epilayers the quasiparticle dynamics is described by the effective
Hamiltonian $\widehat{H}_{D}$, which replaces the Dirac-Weyl Hamiltonian of
ideal graphene. An important effect, apparent in \mbox{Fig.\
  \ref{fig:electronics}}, is the strong resonant interaction of the Dirac
bands with the interface states. This interaction should mediate a
quasiparticle scattering by interface phonons. By artificially increasing the
top layer separation we can trace the transition from an epitaxial to a
freestanding graphene spectrum [cf.\ Fig.\ \ref{fig:electronics}(g),(h)]. 
The band splitting as well as the resonant interaction decrease
sharply, yet, the typical pattern of the resonant coupling is clearly seen in
\mbox{Fig.\ \ref{fig:electronics}(h).}
Moreover, it is apparent in \mbox{Fig.\ \ref{fig:electronics}(h)}, that the resonant
interaction is subject to certain selection rules: while one Dirac branch strongly 
couples with the interface state, the interaction matrix element vanishes for the other.
Hence, the dispersion of only one Dirac band is significantly affected 
at equilibrium separation of the top layer \mbox{[cf. Fig.\ \ref{fig:electronics}(f),(i),(j)]}.
The resonant interaction is much more
pronounced on the Si-face, where the coupling matrix element is about
\unit[50]{meV} [as estimated from \mbox{Fig.\
\ref{fig:electronics}(f)]}. This is in accord with the lower carrier
mobility observed on the Si-face in comparison to the C-face.\cite{Seyller08,Hass08,Orlita08} 
As visible in \mbox{Fig.\ \ref{fig:electronics}}(c),(d), the resonant coupling also occurs
close to the Fermi energy. 

For a $(6\sqrt{3}$$\times$$6\sqrt{3})$R30 SiC($0001$) surface a similar substrate-mediated
interaction of the Dirac bands has to be considered, albeit with a different
translational symmetry. Due to the weak coupling of the graphene layer one expects a
qualitatively similar scenario. Indeed, the gap opening as well as the
formation of flat interface ``midgap states'' were found by \emph{ab initio}
modeling\cite{Kim08} of photoemission spectra for a $(6\sqrt{3}$$\times$$6\sqrt{3})$R30
substrate.

The authors acknowledge financial support by the International Center for
Molecular Materials (ICMM) at the University of Erlangen-N\"urnberg and by the
European Science Foundation (ESF) under the EUROCORES Program EuroGRAPHENE CRP
GRAPHIC-RF (DFG-grant PA 516/8-1).

\end{document}